\documentclass[letter,nofootinbib,superscriptaddress,twocolumn]{revtex4}

\usepackage{amsmath}
\usepackage{amssymb}
\usepackage{braket}
\usepackage{amssymb}
\usepackage{color}
\usepackage{xcolor}
\usepackage{graphicx}
\usepackage{latexsym}
\usepackage{listings}
\usepackage{mathrsfs}
\usepackage{aas_macros}
\usepackage{tikz}

\usepackage{subfigure}

\usepackage{verbatim}
\usepackage{tabularx}
\usepackage{ragged2e}
\usepackage{multirow}
\usepackage{amsbsy}

\usepackage{numprint}
\usepackage{makecell}
\usepackage[utf8]{inputenc}

\usepackage[colorlinks,citecolor=blue]{hyperref}

\newcommand{\paren}[1]{\left( #1 \right)}

\newcommand{\nubhlight}{$\nu\texttt{bhlight}$${}$}

\newcommand{\rhou}{\lfloor\rho\rfloor}

\newcommand{\unsim}{{\mathord{\sim}}}

\npdecimalsign{.}
\npthousandsep{,}

\newcolumntype{Y}{>{\RaggedRight\arraybackslash}X}

\graphicspath{{./figures/}}

\sloppypar

\begin{document}



\title{Full Transport Model of GW170817-Like Disk Produces a Blue
  Kilonova}

\author{Jonah M. Miller}
\email{jonahm@lanl.gov}
\affiliation{CCS-2, Los Alamos National Laboratory, Los Alamos, NM 87545, USA}
\affiliation{Center for Theoretical Astrophysics, Los Alamos National Laboratory, Los Alamos, NM 87545, USA}
\affiliation{Center for Nonlinear Studies, Los Alamos National Laboratory, Los Alamos, NM 87545, USA}

\author{Benjamin R. Ryan}
\affiliation{CCS-2, Los Alamos National Laboratory, Los Alamos, NM 87545, USA}
\affiliation{Center for Theoretical Astrophysics, Los Alamos National Laboratory, Los Alamos, NM 87545, USA}

\author{Joshua C. Dolence}
\affiliation{CCS-2, Los Alamos National Laboratory, Los Alamos, NM 87545, USA}
\affiliation{Center for Theoretical Astrophysics, Los Alamos National Laboratory, Los Alamos, NM 87545, USA}

\author{Adam Burrows}
\affiliation{Department of Astrophysical Sciences, Princeton University, Princeton, NJ, USA}

\author{Christopher J. Fontes}
\affiliation{XCP-5, Los Alamos National Laboratory, Los Alamos, NM 87545, USA}
\affiliation{Center for Theoretical Astrophysics, Los Alamos National Laboratory, Los Alamos, NM 87545, USA}

\author{Christopher L. Fryer}
\affiliation{CCS-2, Los Alamos National Laboratory, Los Alamos, NM 87545, USA}
\affiliation{Center for Theoretical Astrophysics, Los Alamos National Laboratory, Los Alamos, NM 87545, USA}

\author{Oleg Korobkin}
\affiliation{CCS-7, Los Alamos National Laboratory, Los Alamos, NM 87545, USA}
\affiliation{Center for Theoretical Astrophysics, Los Alamos National Laboratory, Los Alamos, NM 87545, USA}

\author{Jonas Lippuner}
\affiliation{CCS-2, Los Alamos National Laboratory, Los Alamos, NM 87545, USA}
\affiliation{Center for Theoretical Astrophysics, Los Alamos National Laboratory, Los Alamos, NM 87545, USA}

\author{Matthew R. Mumpower}
\affiliation{Center for Theoretical Astrophysics, Los Alamos National Laboratory, Los Alamos, NM 87545, USA}
\affiliation{T-2, Los Alamos National Laboratory, Los Alamos, NM 87545, USA}

\author{Ryan T. Wollaeger}
\affiliation{CCS-2, Los Alamos National Laboratory, Los Alamos, NM 87545, USA}
\affiliation{Center for Theoretical Astrophysics, Los Alamos National Laboratory, Los Alamos, NM 87545, USA}

%
%


\begin{abstract}
  The 2017 detection of the inspiral and merger of two neutron stars
  in gravitational waves and gamma rays was accompanied by a
  quickly-reddening transient. Such a transient was predicted to occur
  following a rapid neutron capture (r-process) nucleosynthesis event,
  which synthesizes neutron-rich, radioactive nuclei and can take
  place in both dynamical ejecta and in the wind driven off the
  accretion torus formed after a neutron star merger. We present the
  first three-dimensional general relativistic, full transport
  neutrino radiation magnetohydrodynamics (GRRMHD) simulations of the
  black hole-accretion disk-wind system produced by the GW170817
  merger. We show that the small but non-negligible optical depths
  lead to neutrino transport globally coupling the disk electron
  fraction, which we capture by solving the transport equation with a
  Monte Carlo method. The resulting absorption drives up the electron
  fraction in a structured, continuous outflow, with electron fraction
  as high as $Y_e\unsim 0.4$ in the extreme polar region. We show via
  nuclear reaction network and radiative transfer calculations that
  nucleosynthesis in the disk wind will produce a blue kilonova.
\end{abstract}

\maketitle



\section{Introduction}

In August, 2017, the inspiral and merger of a
pair of neutron stars (GW170817) was jointly detected by gravitational
wave detectors and electromagnetic telescopes around the world
\cite{GW170817PRL}. This detection confirms that such mergers are
central engines of short gamma ray bursts
\cite{GW170817GRB,EichlerBNSGRB,NarayanGRB} and a site of r-process
nucleosynthesis \cite{GW1708MultiMessanger,Rosswog2018Nucleo}, where
the heaviest elements in our universe are formed
\cite{Blinnikov,LattimerCBC1,LattimerCBC2,CoteRProcess}.

The radioactive decay of r-process elements produces an optical and
infra-red afterglow---the \textit{kilonova}
\cite{MetzgerRProcess,CoteRProcess}, which was observed clearly in the
aftermath of GW170817 \cite{GW1708MultiMessanger}. This afterglow is
likely driven by at least two components
\cite{Cowperthwaite_2017,TanvirHST,Tanaka2017KN}: a ``blue'' kilonova
driven by polar outflow \cite{EvansBlueKN,Nicholl_2017} and a ``red''
kilonova driven by equatorial outflow
\cite{Chornok_2017,TanvirHST,TrojaHST}. These distinct components are
believed to arise due to the different compositions of these outflows
\cite{Cowperthwaite_2017,TanvirHST,Tanaka2017KN}. Relatively neutron
rich outflows with an electron fraction $Y_e \lesssim 0.25$ can
produce lanthanides \cite{Mumpower12,Lippuner_2015}, which are opaque
to blue light \cite{Kasen_2013, Fontes15, Fontes19}. Less neutron-rich
outflows ($Y_e \gtrsim 0.25$) will produce nucleosynthetic yields
which allow blue light to escape the photosphere \cite{Lippuner_2015,
  Martin_2015}.

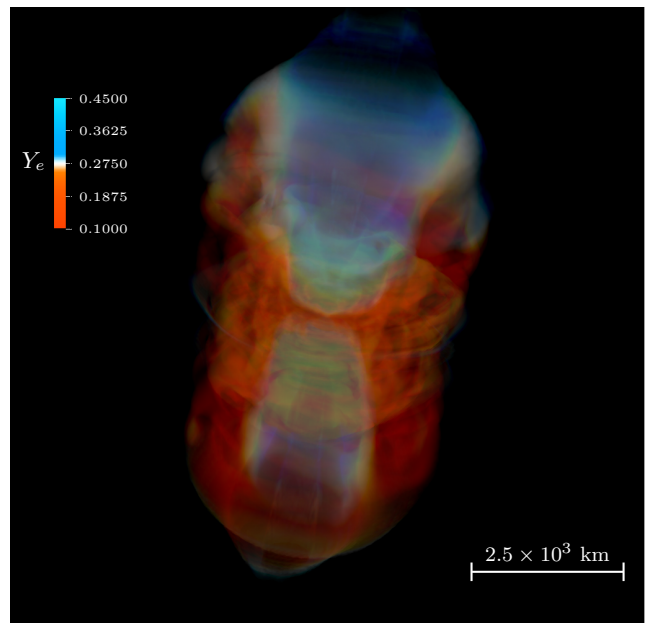
\begin{figure}[t!]
  \resizebox{\columnwidth}{!}{
    \begin{tikzpicture}
      \newcommand{\tcw}{\tiny\color{white}}
      \newcommand{\dcbar}{-0.445}
      \node[inner sep=0pt] (vr) at (0,0)
      {\includegraphics
        [width=\columnwidth,
        clip,trim={0 0 0 0}]
        {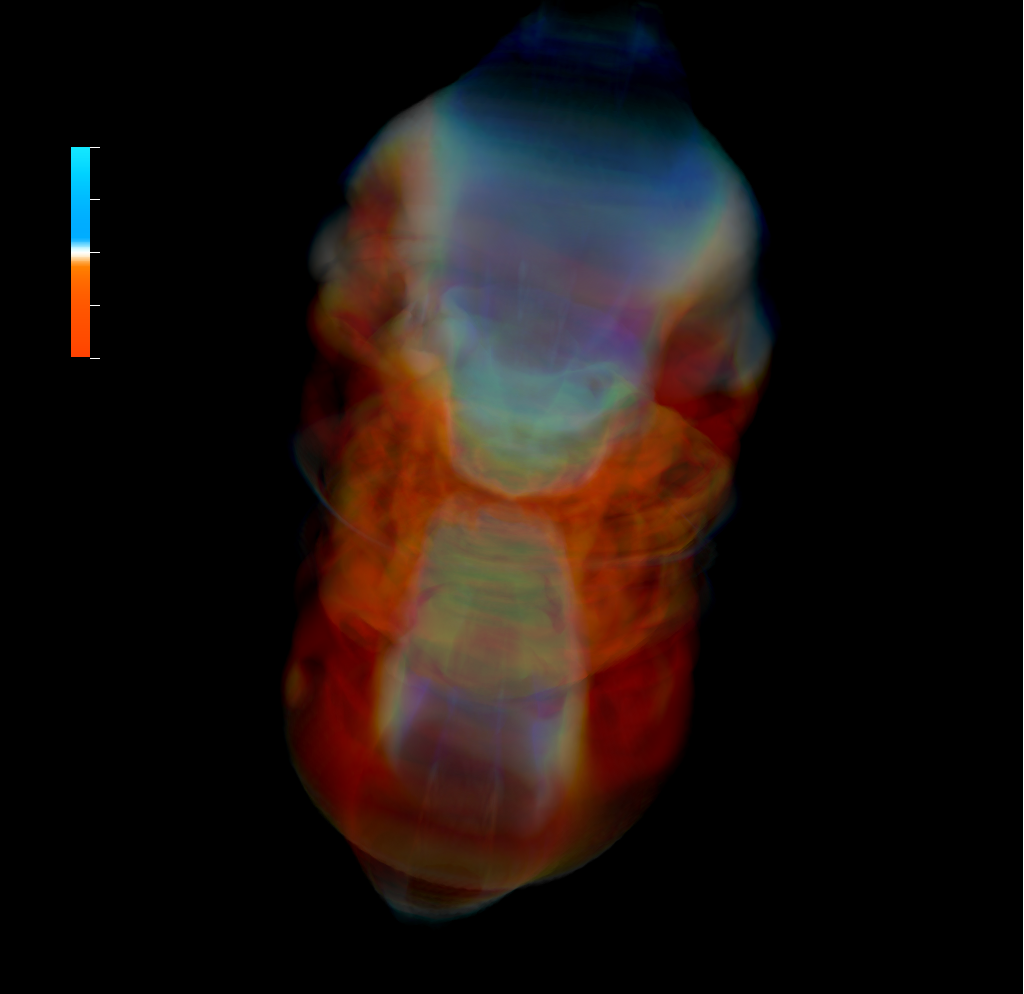}};
      \node at (-4,2.1) {\small \color{white} $Y_e$};
      \draw[|-|,white,thick]
      (1.95,-3.5)
      -- ++(1.05,0) node[above]
      {\footnotesize \color{white} $2.5\times 10^3$ km}
      -- ++(1.05,0);
      \draw
        (-3.5,2.955) node[right] {\tcw 0.4500}
      -- ++(0,\dcbar) node[right] {\tcw 0.3625}
      -- ++(0,\dcbar) node[right] {\tcw 0.2750}
      -- ++(0,\dcbar) node[right] {\tcw 0.1875}
      -- ++(0,\dcbar) node[right] {\tcw 0.1000};
    \end{tikzpicture}
  }
  \caption{Volume rendering of the electron fraction $Y_e$ of material
    in the disk-wind system after $\unsim 31.7$ ms. Opacity in image
    is proportional to temperature.}
  \label{fig:outflow:3d}
\end{figure}

Several mechanisms can produce these outflows
\cite{RosswogReview2015,FernandezReview2016}. Tidal ejecta typically
produce a red component, while shock-driven, near-polar dynamical
ejecta can potentially be blue
\cite{SekiguchiBNS,FoucartBNS2016,RadiceBNS2016,BovardBNS2017,Martin_2018}. Wind
off of a remnant hypermassive, supramassive, or stable neutron star
can also be blue
\cite{Dessart2009nuWind,PeregoNSWind,Martin_2015}. Finally, a
remnant-disk system can drive a wind
\cite{SekiguchiBNS,PeregoNSWind,Ruffert1997DiskWind,PophamNDAF,ShibataBNSDiskMass,Shiabata2007GRRMHD,Surman_2008,MetzgerGRB2008,Beloborodov2008,MetzgerPrioQuataert2009,LeeRR2009,FernandezMetzger2013_2D,FernandezMetzger2014Disk2d,JaniukGRB,JustComprehensive2015,RichersKasen,FoucartPostMerger,Wu2016Disk3rdPeak,LippunerRProcess,NouriPostMerger,SiegelMetzger3DBNS,FernandezLongTermGRMHD,Foucart2018Evaluating}. For
this last source, the composition is as-yet uncertain. Some studies
show the disk wind to have an electron fraction ranging from
$Y_e\unsim 0.2 - 0.4$ and thus produce a blue component
\cite{FernandezMetzger2013_2D,FernandezMetzger2014Disk2d,
  JaniukGRB,JustComprehensive2015,FernandezLongTermGRMHD,Foucart2018Evaluating}.
Other work shows the disk wind to be uniformly composed of
$Y_e\unsim 0.2$ material that produces only a red component
\cite{Wu2016Disk3rdPeak,SiegelMetzger3DBNS}.

We focus on the evolution of the post-merger disk. Until now, studies
of the remnant disk wind have employed various approximations to the
neutrino transport, neutrino-matter coupling, or magnetohydrodynamics
(MHD). In this work, we present, for the first time, fully
three-dimensional general-relativistic radiation magnetohydrodynamics
(GRRMHD) simulations of a post-merger disk system with full neutrino
transport using a Monte Carlo method.

We model a black hole accretion disk system which may have formed from
the GW170817 merger \cite{Shibata170817}. Magnetohydrodynamic turbulence
\cite{BalbusHawley91} drives a wind
\cite{BlandfordPayneHydromagneticOutflows} off the disk. We find the
electron fraction of this outflow ranges from $Y_e\unsim 0.2$ to
$Y_e\unsim 0.4$. Moreover, we find that the composition of the outflow
varies significantly with angle off of the midplane, suggesting that
the observed character of the outflow depends heavily on viewing
angle. Thus, a blue, wind-produced kilonova will be visible if the
remnant is viewed close to the polar axis.



\begin{figure}[tb!]
  \includegraphics[width=\columnwidth]{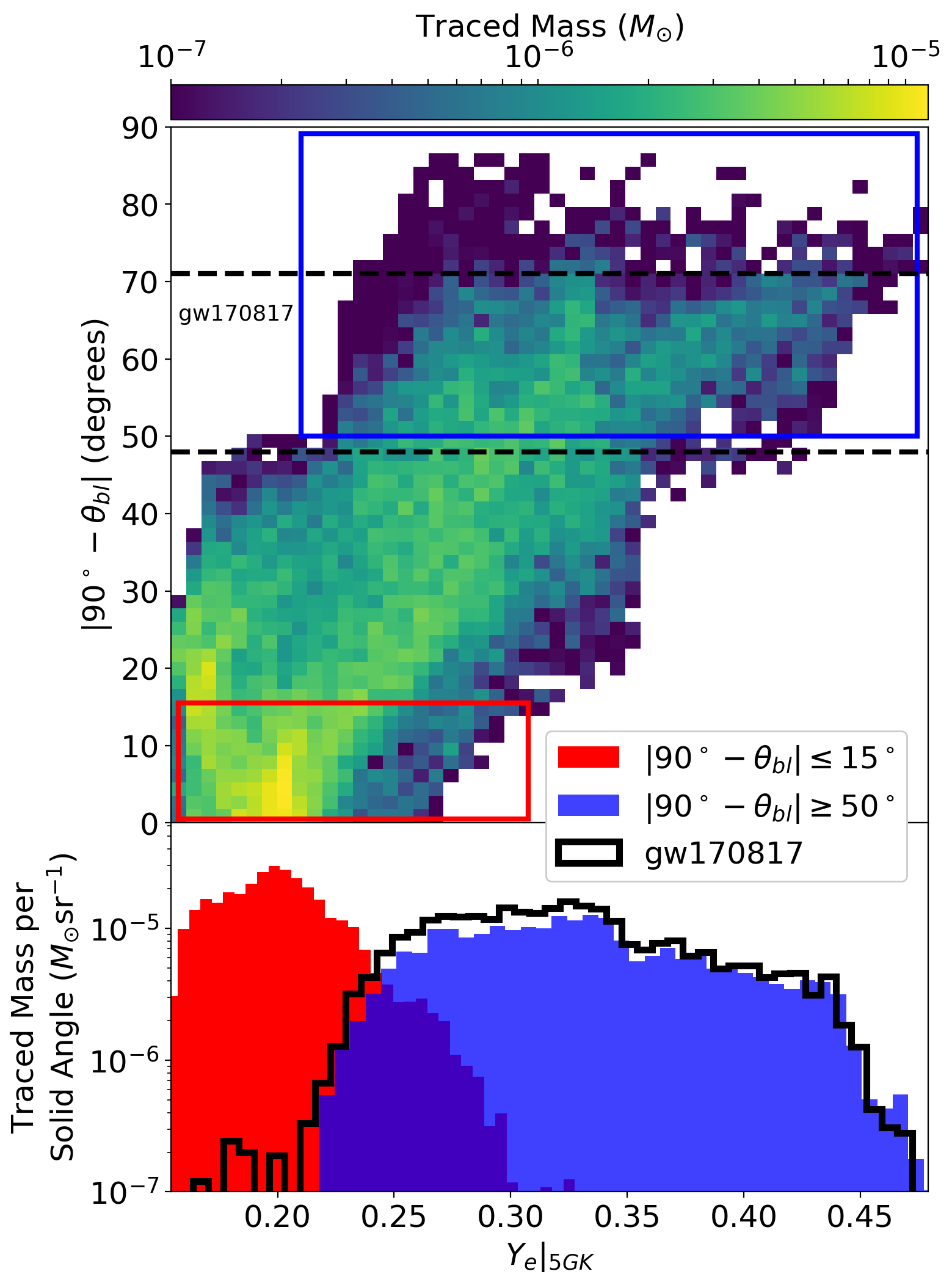}
  \caption{Top: Electron fraction of gravitationally unbound material
    at 5 GK vs. latitude, $|90-\theta_{\text{bl}}|$. Boxes represent
    cuts through the data. Red is neutron-rich, blue is
    neutron-poor. Black dashed lines represent approximate bounds on
    viewing angle for gw170817, as given by
    \cite{LIGOAngleBiwer}. (Although angle matters, an observation
    integrates over many lines of sight.) Bottom: Distribution per
    solid angle of electron fraction in material in boxed regions.}
  \label{fig:Ye:v:theta}
\end{figure}

\begin{figure}[tb]
  \includegraphics[width=\columnwidth]{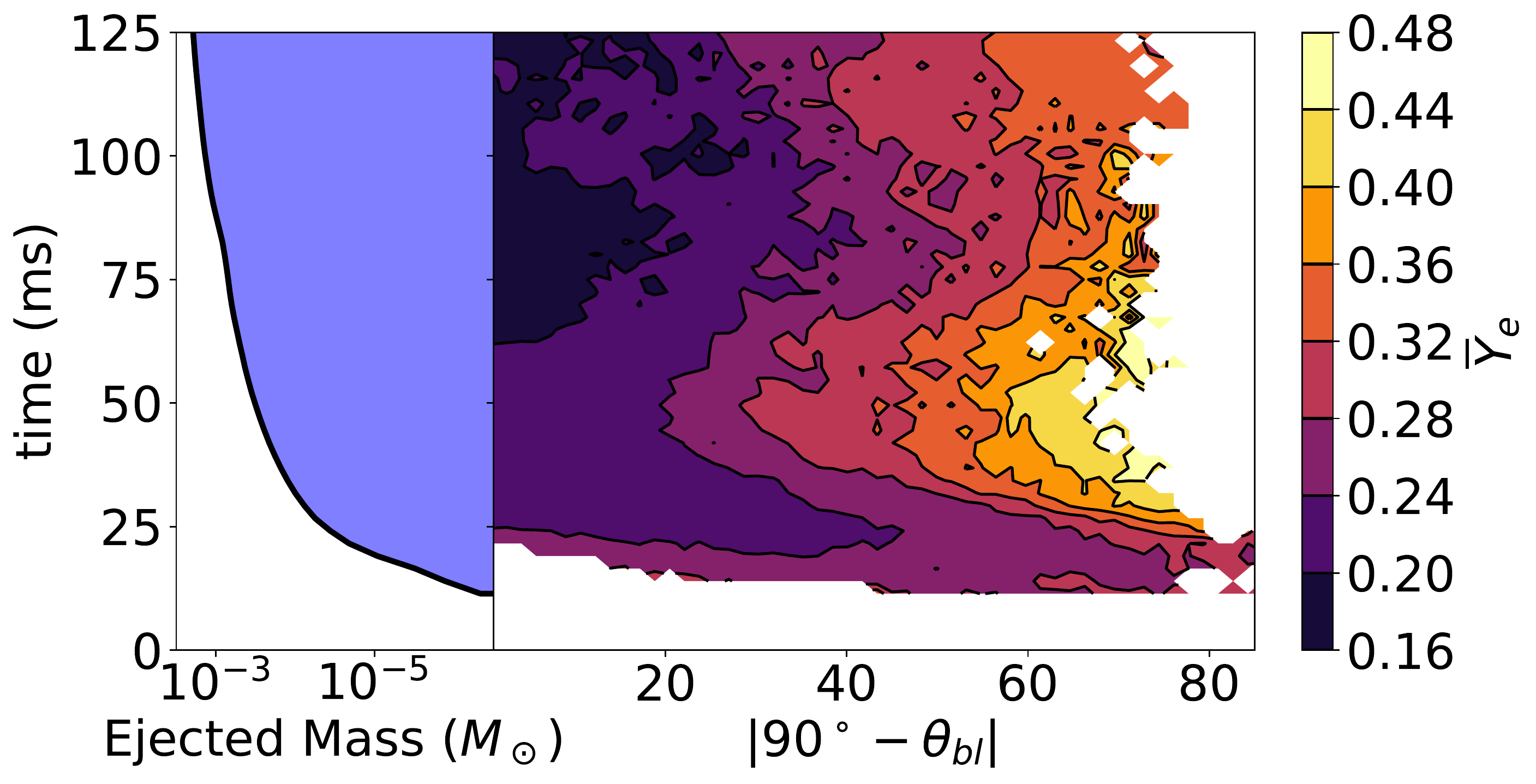}
  \caption{Left: Total mass in the outflow as a function of
    time. Right: Average electron fraction $Y_e$ of gravitationally
    unbound material at an extraction radius of $r\sim 10^3$ km as a
    function of latitude and time.}
  \label{fig:ye-spacetime}
\end{figure}

\section{Methods}

We perform a GRRMHD simulation in full three
dimensions with our code, \nubhlight \cite{nubhlight}. We assume a
Kerr background metric, consistent with the relatively small disk mass
compared to black hole mass. The radiation transport is treated via
explicit Monte Carlo and the MHD is treated via finite volumes with
constrained transport. The two methods are coupled via operator
splitting.

We use the SFHo equation of state \cite{SFHoEOS} as
tabulated in \cite{stellarcollapsetables,stellarcollapseweb} and the
neutrino-matter interactions described in \cite{nubhlight} and
tabulated in \cite{BurrowsNeutrinos}. For initial data, we use
parameters consistent with a remnant from GW170817
\cite{Shibata170817,GW170817PRL,GW170817PropertiesPRX}: an equilibrium
torus \cite{FishboneMoncrief} of mass $M_d = 0.12$ $M_\odot$ and
constant electron fraction $Y_e = 0.1$ around a black hole of mass
$M_{BH} = 2.58$ $M_\odot$ and dimensionless spin $a = 0.69$. We thread
our torus with a single poloidal magnetic field loop such that the
minimum ratio of gas to magnetic pressure is 100.



\begin{figure}[tb]
  \includegraphics[width=\columnwidth]{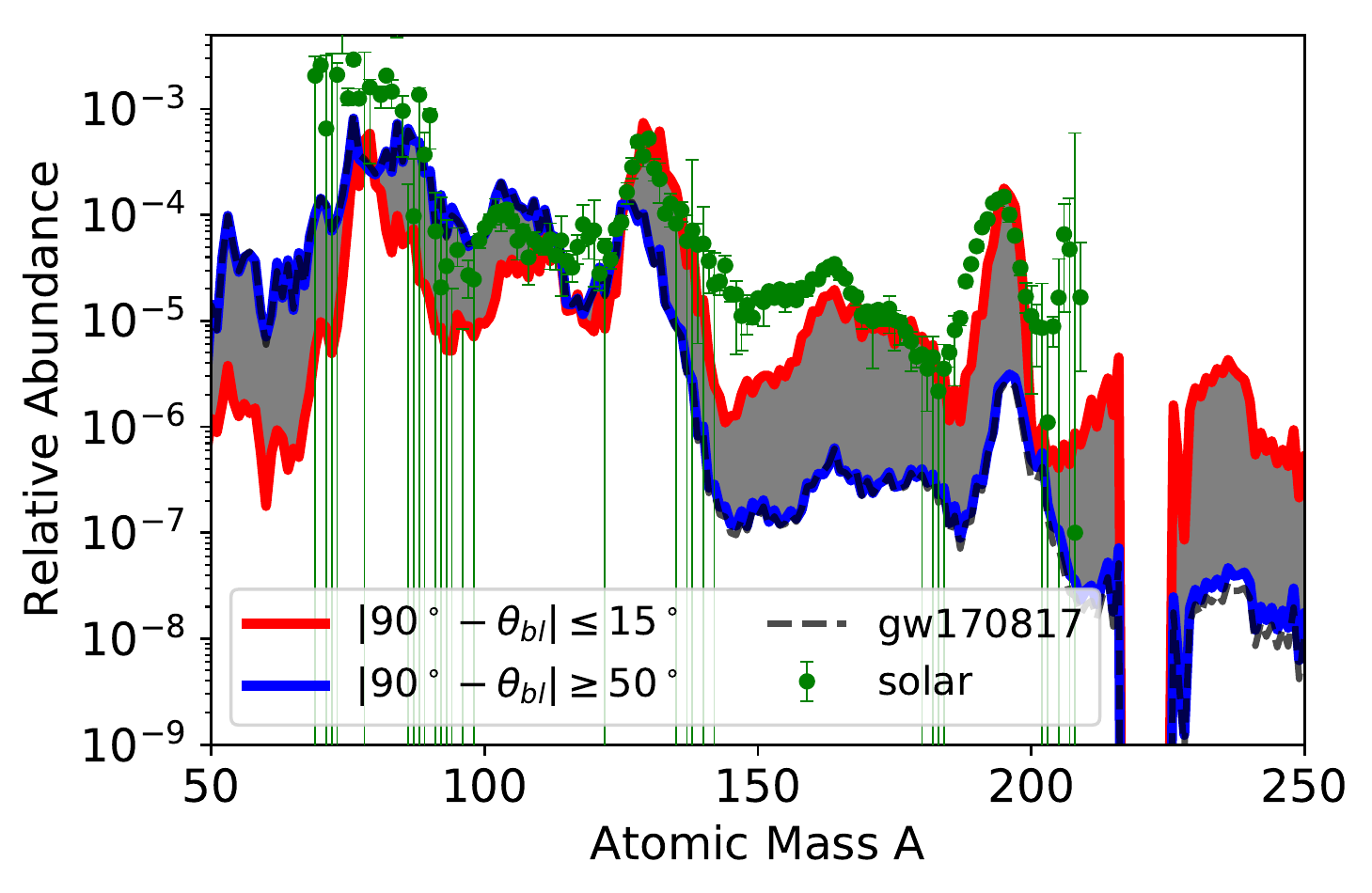}
  \caption{Relative abundance of yields for disk outflow: red for
    material with $<15^\circ$ off the midplane and blue for material
    $>50^\circ$. Gray shading shows the range of values that can be
    attained at intermediate angles. Black dashed line shows yields
    attained in the GW170817 box in figure
    \ref{fig:Ye:v:theta}. Curves are normalized by mass
    fraction. Solar abundances from \cite{Arnould+07} shown in green.}
  \label{fig:yields}
\end{figure}


\section{Outflow Properties}

Our disk drives a wind consistent with other GRMHD simulations of
post-merger disks
\cite{FernandezMetzger2013_2D,FernandezMetzger2014Disk2d,
  JaniukGRB,JustComprehensive2015,FernandezLongTermGRMHD,Foucart2018Evaluating,Wu2016Disk3rdPeak,SiegelMetzger3DBNS},
which expands outward from the disk in polar lobes as shown in figure
\ref{fig:outflow:3d}.  We record material crossing a sphere of radius
$r \sim 10^3$ km. Figure \ref{fig:Ye:v:theta} bins outflow material in
both electron fraction $Y_e$ and in angle off the equator,
$|90^\circ -\theta_{\text{bl}}|$ for Boyer-Lindquist angle
$\theta_{\text{bl}}$, integrated in time. The 90\% confidence interval
for the viewing angle for GW170817 \cite{LIGOAngleBiwer} is bounded by
the dashed lines.

We choose two regions, one close to the midplane, and one far from it,
highlighted in the red and blue rectangles. We bin the electron
fraction in these regions in the red and blue histograms. Regardless
of electron fraction, ejected material has an average entropy, $s$, of
about 20 $k_b/$baryon and an average radial velocity (as measured at a
radius of 1000 km) of about 0.1$c$.

The electron fraction depends on angle off of the midplane and this
dependence persists through time. The right panel of figure
\ref{fig:ye-spacetime} shows the average electron fraction of gravitationally
unbound material passing through a surface at $t\sim 10^3$ km as a
function of angle off the equator and time. For any given time, larger $|90^\circ - \theta_{\text{bl}}|$ correlates with larger $Y_e$.

\begin{figure}[tb!]
  \includegraphics[width=\columnwidth]{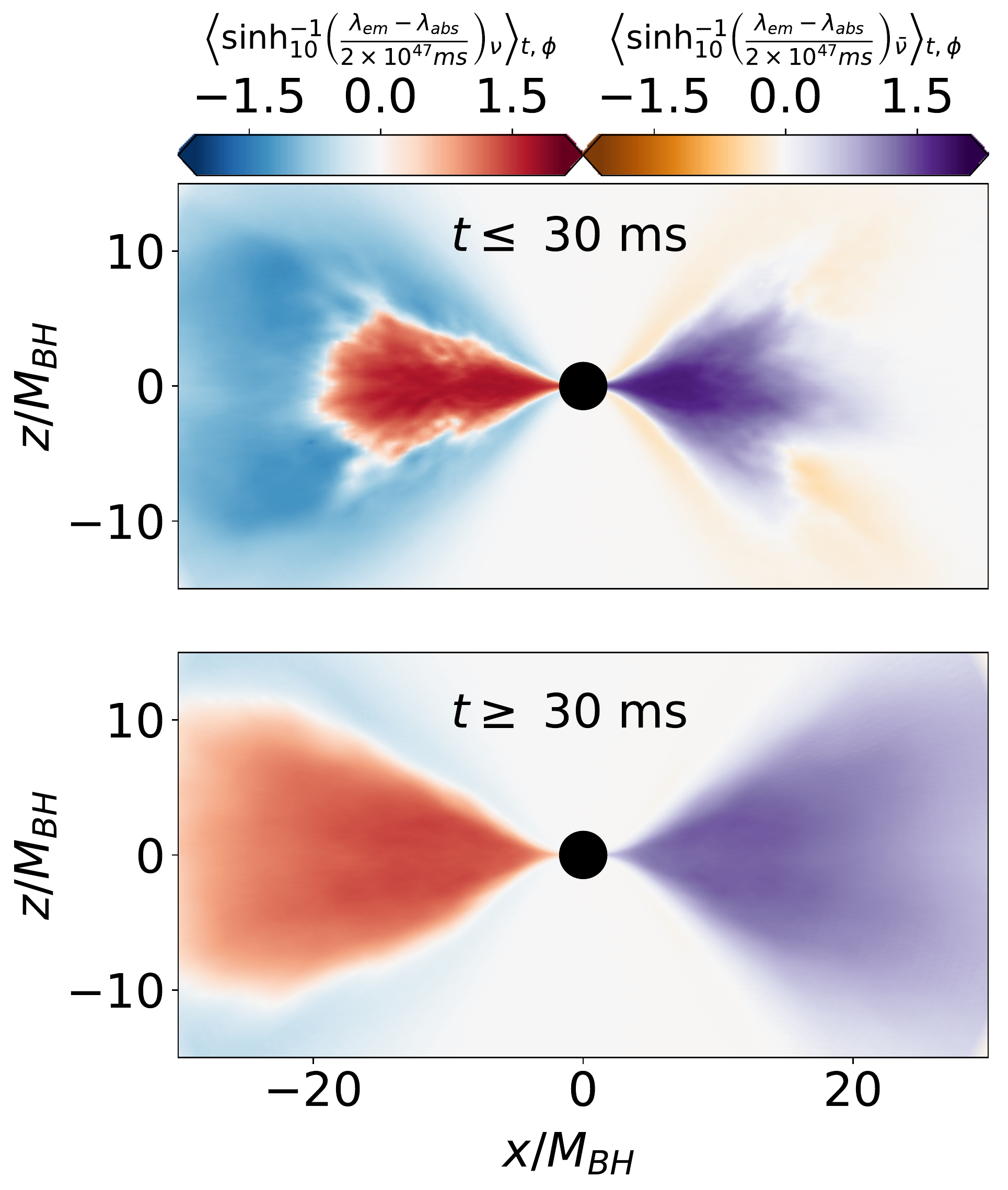}
  \caption{Rate of emitted neutrinos minus the rate of absorbed
    neutrinos for electron neutrinos ($x<0$) and electron
    antineutrinos ($x>0$). Averaged over azimuthal angle $\phi$ and in
    time from 0 to 30 ms (top) and from 30 ms to 127 ms (bottom).}
  \label{fig:sinh:rates}
\end{figure}

We use the nuclear reaction network SkyNet \cite{Skynet} to compute
nucleosynthetic yields on tracer particles advected with
gravitationally unbound material. We start the network calculation
when the tracer reaches $T \unsim 10$~GK and we assume a nuclear
statistical equilibrium (NSE) composition at that time. The network is
run up to $t = 10^9$~s assuming homologous expansion
($\rho \propto t^{-3}$) and uses the same nuclear physics inputs as in
\cite{LippunerRProcess, Roberts_2017}, namely: 8000 nuclides and
140,000 nuclear reactions, including fission, with rates from
\cite{REACLIB, frankel:47, panov:10, mamdouh:01, wahl:02, fuller:82,
  oda:94, langanke:00}.

Figure \ref{fig:yields} plots nucleosynthetic yields. We plot three
angular cuts: in red for material near the midplane, in blue for
material near the poles, and in black for material within the viewing
angle for GW170817 \cite{LIGOAngleBiwer}. We sketch out the range of
possible yields in gray. The second, rare-earth and third peaks are
suppressed by up to a factor of 100 with respect to the first peak in
the polar regions.



\section{Outflow Mass}

The left panel of figure \ref{fig:ye-spacetime} shows the total mass
in the outflow as a function of time. Due to computational cost, we
did not run our simulation for long enough to observe the total amount
of mass that becomes gravitationally unbound. As a lower bound, we
report the amount of material with Bernoulli parameter $B_e > 0 $
\cite{NovikovThorne} at a radius greater than 125 gravitational radii
($\unsim 500$ km) at the end of the simulation ($\unsim 127$
ms). (This includes material that has already left the domain.) We
find this to be about $4.33\times 10^{-3} M_{\odot}$ and the ratio of
mass in the outflow to accreted mass is about 9\% by this time in the
simulation. About 18\% of this outflow has an electron fraction of
$Y_e \geq 0.275$ and about 14.5\% is within the expected range of
viewing angles for GW170817.

\section{Neutrino Transport}

A characterization of the importance of neutrino absorption is the
neutrino absorption optical depth $\tau$ of the disk. $\tau \ll 1$
implies free-streaming and $\tau \gg 1$ implies no neutrino can
escape. At relatively early times ($t \lesssim 30$ ms), we find
$\tau \sim 10$. In this phase, $Y_e$ evolution is dominated by
emission of electron neutrinos in the core of the disk and their
absorption in the corona. At later times ($t \gtrsim 30$ ms), the disk
achieves a quasistationary state with $\tau \sim 0.1$. Although this
later stage is emission dominated, reaching it requires properly
treating absorption.

\begin{figure}[tb!]
  \includegraphics[width=\columnwidth]{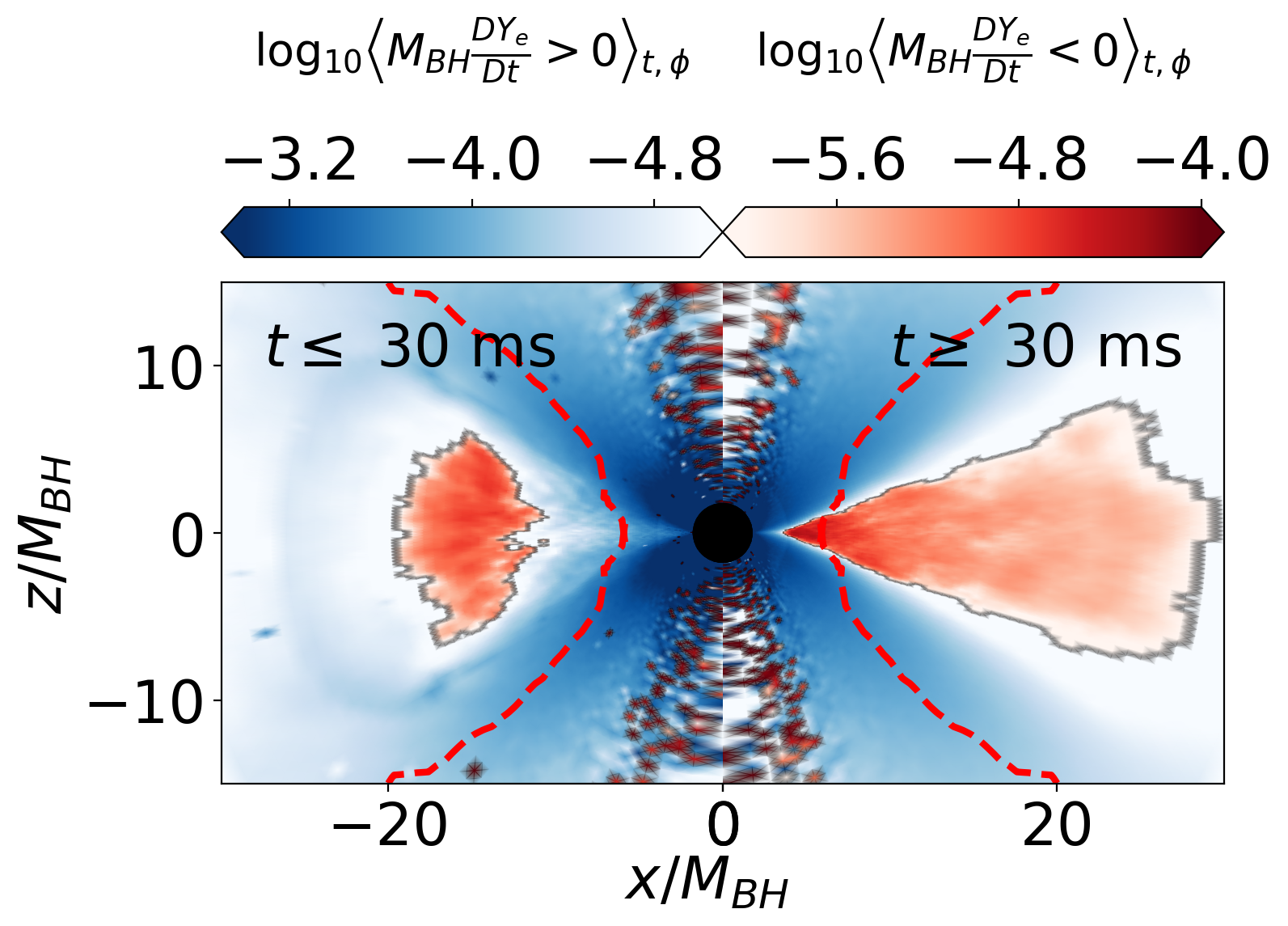}
  \caption{Lagrangian derivative of electron fraction due to emission
    or absorption of neutrinos: blue for an increase in $Y_e$ and red
    for a decrease. Averaged over azimuthal angle $\phi$ and in time
    from 0 to 30 ms (left) and from 30 ms to 127 ms (right). Red
    curves define a surface at which gravitationally unbound material
    reaches within 5\% of its asymptotic $Y_e$ at infinity, roughly
    indicating where neutrino interactions significantly effect the
    electron fraction of escaping material. Very little material
    becomes unbound closer to the black hole than the innermost radius
    of the red curves.}
  \label{fig:dy:dt}
\end{figure}

Figure \ref{fig:sinh:rates} shows this transition. We plot for both
phases the $\sinh_{10}^{-1}$ of the rate
$\lambda = \partial N/\partial t$ of emitted neutrinos minus the rate
of absorbed neutrinos for electron neutrinos ($x<0$) and electron
antineutrinos ($x>0$), where we define $\sinh^{-1}_{10}$ as the
inverse of $\sinh_{10}(x) := (10^x - 10^{-x})/2$ such that for
$|x| \gtrsim 10$,
$\sinh^{-1}_{10}(x)\to \text{sign}(x)\log_{10}(|2x|)$.  Red and orange
imply $Y_e$ is decreasing due to neutrino interactions. Blue and
purple imply it is increasing.  Figure \ref{fig:dy:dt} shows the
resulting change in the electron fraction in the Lagrangian frame for
each phase.



\section{Electromagnetic Counterpart}

We compute spectra from the kilonova assuming spherically symmetric
outflow composed of nucleosynthetic yields produced in material with
$|90^\circ - \theta_{\text{bl}}| \leq 15^\circ$ and
$|90^\circ - \theta_{\text{bl}}| \geq 50^\circ$. For comparison, we
compute spectra for an outflow with solar-like abundances such as
those reported in \cite{SiegelMetzger3DBNS}. For the former, we assume
an outflow mass of $M_e = 10^{-2}$ $M_\odot$, consistent with our
results. For the latter, we assume an outflow mass of
$M_e = 2\times 10^{-2}$ $M_\odot$, consistent with
\cite{SiegelMetzger3DBNS}. We use a mean radial velocity of $0.1c$.

To compute spectra for each model, we simulate radiative transfer with
the Monte Carlo code {\tt SuperNu} \cite{Wollaeger14,Wollaeger19},
using opacity from the LANL suite of atomic physics codes
\citep{Fontes15}. We use a complete suite of lanthanide opacities
\citep{Fontes19}, and some representative wind opacities
\citep{WollaegerEjectaMorphology}. These calculations do not
  explore the effect of aspherical morphology or uncertainties in
  r-process heating or composition.

\begin{figure}[tb!]
  \includegraphics[width=\columnwidth]{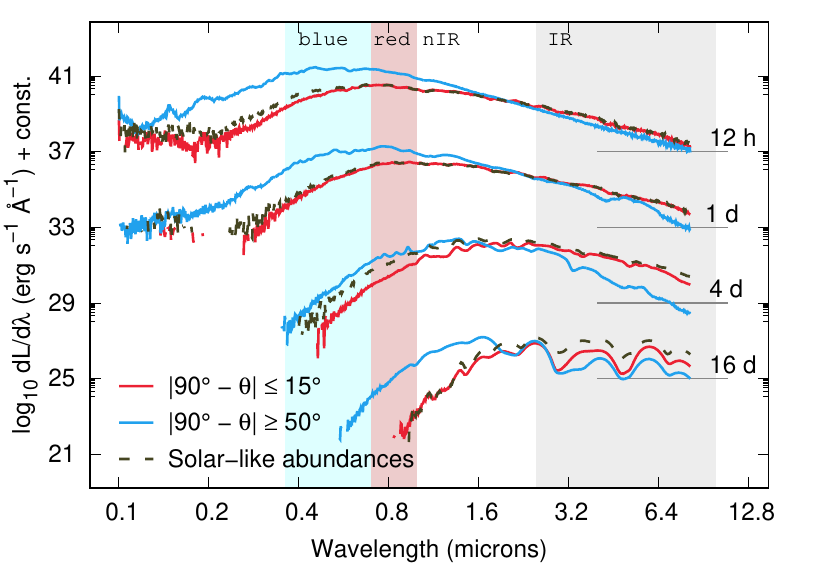}
  \caption{Electromagnetic spectra for spherically symmetric outflow
    composed of nucleosynthetic yields produced in material
    $<15^\circ$ off the midplane, $>50^\circ$ degrees off the
    midplane, and of solar abundances such as those produced in tidal
    ejecta or outflows like those reported in
    \cite{SiegelMetzger3DBNS}. At 5000\AA, the polar outflow is
    $\unsim 12\times$ more luminous than the more neutron-rich
    outflows.}
  \label{fig:LC}
\end{figure}

Figure \ref{fig:LC} shows computed spectra for several epochs after
merger. At early times, the polar outflow produces more luminous
spectra peaked at a blue wavelength, consistent with a blue kilonova.
Differences in these early-time spectra amount to about a 2 magnitude
difference in brightness between polar and equatorial outflows. At
late times, the more neutron-rich outflows are more luminous and
peaked at long wavelengths, consistent with a red kilonova. The
luminosity peaks at $\unsim 3\times 10^{41}$ erg$/$s after
$\unsim 0.3$ days for the
$|90^\circ - \theta_{\text{bl}}| \leq 15^\circ$ outflow, at
$3\times 10^{40}$ erg$/$s after $\unsim 2$ days for the
$|90^\circ - \theta_{\text{bl}}| \geq 50^\circ$ outflow, and
$4\times 10^{40}$ erg$/$s after $\unsim 4$ days for the solar-like
outflow.

\section{Outlook}

We have explored a possible disk-driven outflow
from the remnant of the GW170817 merger using the first full transport
GRRMHD simulations of a post-merger accretion disk system. We
calculate nucleosynthetic yields and spectra of the electromagnetic
counterpart that would be observed given these yields. These spectra
indicate a blue kilonova, as viewed off the midplane, and a red one,
as viewed from the midplane We find about 9\% of accreted mass ends up
in this outflow.

The range of electron fractions in our outflow is consistent with
\cite{FernandezMetzger2013_2D,FernandezMetzger2014Disk2d,JustComprehensive2015}
and disagrees with \cite{Wu2016Disk3rdPeak,SiegelMetzger3DBNS}. The
former include neutrino absorption but approximate magnetic fields
with a viscous prescription, while \cite{SiegelMetzger3DBNS} uses
GRMHD but includes neutrino absorption only approximately in
post-processing. Our full treatment allows us to conclude that
neutrino absorption is critical to attaining this range of $Y_e$'s.

The electromagnetic counterpart we compute is incomplete, as it is
sourced only by the disk outflow. Depending on the equation of state,
it is possible that a transient remnant neutron star supported by
differential or rigid rotation existed for some time before collapse
to a black hole. For some systems, a long-lived neutron star may be
the remnant \cite{RosswogReview2015,FernandezReview2016}. This remnant
can produce its own separate, potentially massive outflow
\cite{FryerFallback1,FryerFallback2,FryerFallback3}. This outflow and
shock-driven dynamical ejecta
\cite{SekiguchiBNS,FoucartBNS2016,RadiceBNS2016,BovardBNS2017,Martin_2018}
can also contribute to a blue kilonova \cite{PeregoNSWind,
  LippunerRProcess, Martin_2015}.

For a generic binary neutron star merger, our results imply that a
blue kilonova does not necessarily require a short- or long-lived
neutron star remnant. However, these additional sources may be
required to explain both the total mass and the velocity of the source
of the blue component of the afterglow of GW170817 as determined by
early light curve models
\cite{EvansBlueKN,Cowperthwaite_2017,Nicholl_2017,TanvirHST,Tanaka2017KN,TrojaHST}.
Combining our results with other potential sources for a blue
component and the red kilonova from tidal ejecta suggests a three
(or more) component kilonova model, such as the ones described in
\cite{Kasliwal2017concordant,PeregoRadiceBernuzziMultiComponentKN}.


In a black hole-neutron star merger, only the tidal ejecta and
accretion disk are present. An important observational implication of
our model is that this disk-wind system is sufficient to produce a
blue kilonova. This is in contrast to
\cite{Wu2016Disk3rdPeak,SiegelMetzger3DBNS}, which would imply that
black hole-neutron star mergers only produce a red kilonova.

Another important implication of our model is that accurately
capturing the early transient phase of the disk, when optical depths
are relatively large, is critical to correctly predicting the
long-term outflow. Unfortunately, initial conditions are a source of
uncertainty in kilonova disk modeling. A hot hypermassive or
supramassive neutron star can emit its own neutrino flux, which can
reset the electron fraction of the disk. Even in the absence of a hot
remnant, the seed magnetic field is uncertain in both strength and
topology. As the community moves forward more attention should be paid
to both the initial transient phase of the disk and the initial
conditions that drive this early phase.




\section{Acknowledgements}

We thank Francois Foucart, Daniel Siegel,
Ingo Tews, Patrick Mullen, Roseanne Cheng, and especially Ben Prather
for their insight. We also thank our anonymous reviewer for their
thoughtful questions and suggestions.

We gratefully acknowledge support from the U.S. Department of Energy
(DOE) Office of Science and the Office of Advanced Scientific
Computing Research via the SciDAC4 program and Grant DE-SC0018297,
from the U.S. NSF grant AST-174267, and from the U.S. DOE through Los
Alamos National Laboratory (LANL). This work used resources provided
by the LANL Institutional Computing Program. Additional funding was
provided by the LDRD Program and the Center for Nonlinear Studies at
LANL under project number 20170508DR. LANL is operated by Triad
National Security, LLC, for the National Nuclear Security
Administration of the U.S. DOE (Contract
No. 89233218CNA000001). Authorized for unlimited release under
LA-UR-19-22623.

\appendix








\section{Resolution and grid}


We use a radially logarithmic,
quasi-spherical grid in horizon penetrating coordinates, as first
introduced in \cite{HARM}, with
$N_r\times N_\theta \times N_\phi = 192\times 128\times 64$ zones out
to a radius of $\unsim 4\times 10^3$ km. Our grid focuses resolution
at the midplane and we are roughly three times more resolved at the
midplane than a grid with uniform resolution in $\theta$ for an
effective resolution of $N_\theta^{\text{eff}}\sim 384$.

In the region where opacities are nonvanishing (roughly 125 km), we
use more than 38 million Monte Carlo radiation packets at every time
step, resulting in an average packet density of more than 20 packets
per finite volume cell. We track Lagrangian fluid elements via more
than $10^6$ tracer particles, of which about 10\% end up
gravitationally unbound by the final time of $\unsim 127$ ms. We
initialize our tracers so that, at the initial time, they roughly
uniformly sample non-atmosphere regions by volume, as described in
\cite{nubhlight} and suggested in \cite{BovardTracersBNS}.

The focusing effect of our grid reduces the number of grid points
required to resolve the magneto-rotational instability (MRI)
\cite{BalbusHawley91}. Following the definition in \cite{Sano2004}, we
define a quality factor
\begin{equation}
  \label{eq:q:theta}
  Q^{(\theta)}_{\text{mri}} = \frac{2\pi b^{(\theta)}}{\Delta x^{(\theta)}\sqrt{w + b^2}\Omega},
\end{equation}
for the MRI to be the number of grid points per minimum unstable MRI
wavelength inside the disk. Here $b^{(\theta)}$ is the
$\theta$-component of the magnetic field four-vector,
$\Delta x^{(\theta)}$ is grid spacing in the $\theta$ direction, $w$
is the enthalpy of the fluid, and $b^2 = b^\mu b_\mu$ is total
magnetic field strength. Following \cite{SiegelMetzger3DBNS}, we also
define
\begin{equation}
  \label{eq:q:c}
  Q^{(\text{c})}_{\text{mri}} = \frac{b}{b^{(\theta)}} Q^{(\theta)}_{\text{mri}}
\end{equation}
for $b=\sqrt{b^\mu b_\mu}$, which uses the strength of the magnetic
field in the comoving frame, rather than the lab
$\theta$-direction. Unfortunately, the nuclear equation of state we
use \cite{SFHoEOS} makes our enthalpy larger than for an equivalent
disk with an ideal gas equation of state \cite{EHTComparison}, and
this larger enthalpy makes resolving the MRI more challenging.

\begin{figure}[t!b!]
  \includegraphics[width=\columnwidth]{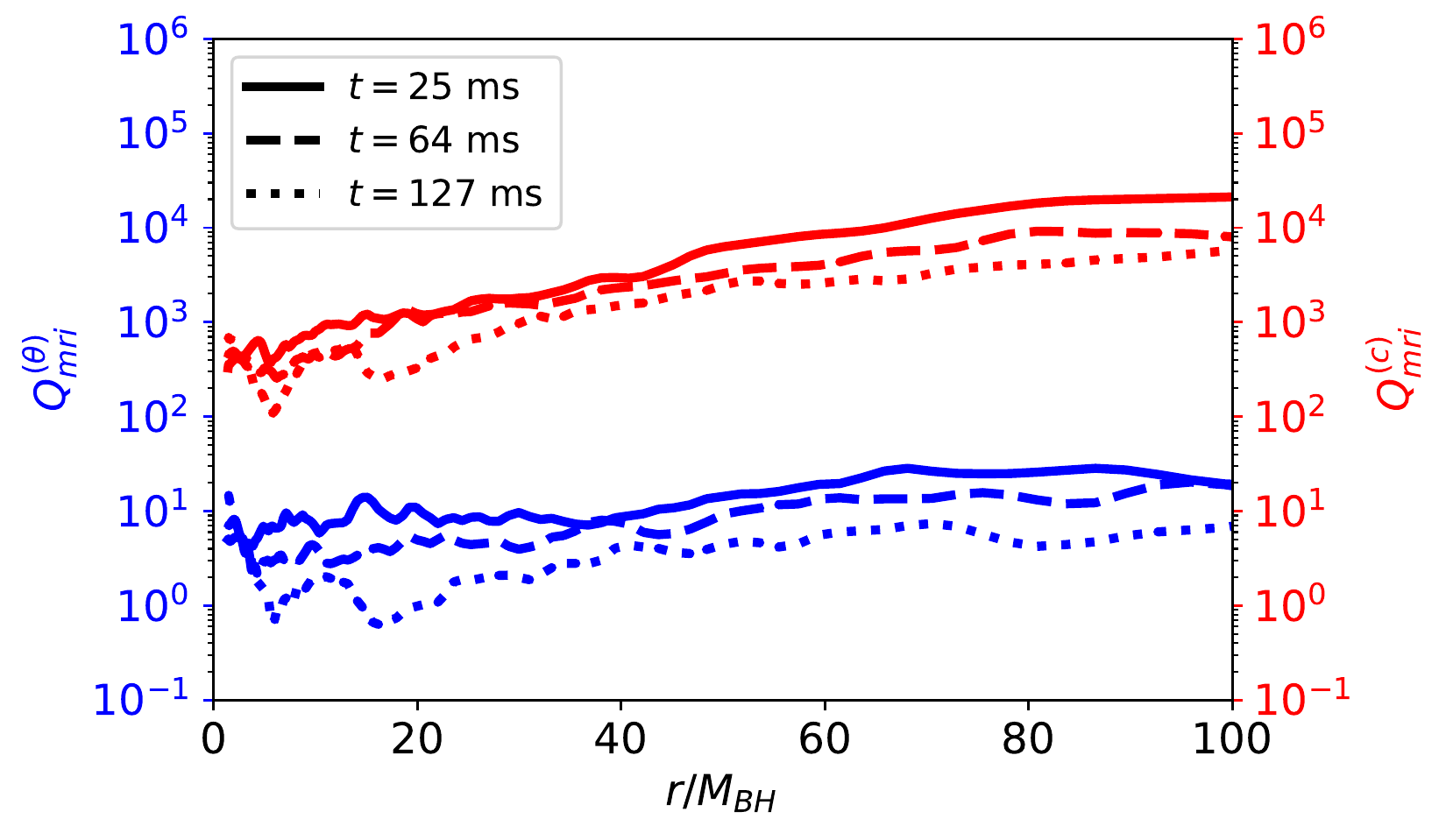}
  \caption{$\phi$-averaged quality factor for the MRI in the midplane
    for three different times. Blue shows the quality factor for
    lab-frame vertical component. Red shows it for the co-moving
    magnetic field.}
  \label{fig:Qmri}
\end{figure}

We plot both $Q_{\text{mri}}^{(\theta)}$ and $Q_{\text{mri}}^{(c)}$ in
the mid-plane averaged over $\phi$ for several times in figure
\ref{fig:Qmri}. On average, we find
$Q^{(\text{c})}_{\text{mri}} \gg 10$ for all time. At early times,
$Q^{(\theta)}_{\text{mri}} \gtrsim 10$. As the disk evolves, this quality
factor drops on average to a minimum of about
$Q^\theta_{\text{mri}} \gtrsim 2$ at $t = 10^4 G M_{BH}/c^3 = 127$ ms. We
note that high-order spatial reconstructions such as the WENO-5
\cite{sashaWENO5} reconstruction we use may effectively improve this
quality factor \cite{EHTComparison}. For various technical issues
related to resolving the MRI in global simulations, see
\cite{EHTComparison}.

\section{Resolution in the radiation sector}

We define the Monte Carlo quality factor
\begin{equation}
  \label{eq:quality:factor}
  Q_{\text{rad}} = \min_{\Omega}\paren{\frac{\partial N}{\partial t} \frac{u}{J}},
\end{equation}
minimized over the simulation domain $\Omega$. $N$ is the number of
emitted Monte Carlo packets, $u$ is gas internal energy density by
volume, and $J$ is the total frequency and angle integrated neutrino
emissivity. $Q_{\text{rad}}$ roughly encodes how well resolved the radiation field
is, with $Q_{\text{rad}}=1$ a marginal value. In our simulation, we find
$Q_{\text{rad}} \geq 100$.

\section{Artificial Atmosphere Treatment}

Since our Eulerian code cannot handle vacuum, we demand densities
$\rho$ obey
\begin{equation}
  \label{eq:floor}
  \rho \geq \text{max}\paren{\frac{\rho_0}{\rhou}\frac{c^4}{G^2M_{BH}^2}\frac{1}{r^2},\rho_{\text{min}}},
\end{equation}
where $\rho_0=10^{-5}$ is a unitless, simulation-dependent parameter,
$\rhou = 1.1\times 10^{13}$ g$/$cm$^3$ is the code unit for density
and $\rho_{\text{min}} = 1.6\times 10^2$ g$/$cm$^3$ is the minimum density in
our tabulated equation of state. We set our initial atmosphere regions
to nearly virial temperatures to prevent the atmosphere from falling
back onto the disk. We track our artificial atmosphere with a passive
scalar and ensure it does not contribute to any reported quantities
such as outflow mass and electron fraction.

As the disk-wind system evolves, outflow will displace the artificial
atmosphere by pushing it through the outer boundary of the
domain. There is no artificial atmosphere remaining after about 63
ms. The total amount of atmosphere displaced in this way before the
wind completely displaces it is roughly $10^{-5} M_{\odot}$, or
$\unsim 10^{-4}$ the mass of the disk and $\unsim 10^{-2}$ the mass of the
outflow. We find that the radial momentum flux in atmosphere regions
is always more than three orders of magnitude less than in wind
regions, giving us confidence that our artificial atmosphere does not
interfere with the dynamics of the outflow.



\bibliography{nubhlight}

\end{document}